\documentclass[12pt]{article}

\usepackage{newtxtext,newtxmath}

\usepackage{graphicx}

\usepackage[letterpaper,margin=1in]{geometry}

\usepackage[overload]{empheq}
\usepackage{cases} 

\renewenvironment{abstract}
	{\quotation}
	{\endquotation}

\date{}

\makeatletter
\renewcommand{\fnum@figure}{\textbf{Figure \thefigure}}
\renewcommand{\fnum@table}{\textbf{Table \thetable}}
\makeatother

\usepackage{scicite}

\usepackage{url}

\def\scititle{
	Associative and Segregative \\Liquid-Liquid Phase Separation in \\Macromolecular Solutions
}
\title{\bfseries \boldmath \scititle}

\author{
	Remco~Tuinier$^{{1}\ast}$,
	{\'A}lvaro~Gonz{\'a}lez~Garc{\'i}a$^{1,2}$\and
	\small$^{1}$Laboratory of Physical Chemistry, Department of Chemical Engineering and Chemistry,\and
    \small and Institute for Complex Molecular Systems (ICMS) \and
    \small Eindhoven University of Technology, P.O. Box 513, 5600 MB, Eindhoven, The Netherlands.\and
    \small$^{2}$International School of Utrecht, 3528 AG, Utrecht, The Netherlands.\and
	\small$^\ast$Corresponding author. Email: r.tuinier@tue.nl\and
}

\begin{document} 

\maketitle

\clearpage
\begin{abstract} \bfseries \boldmath
We investigate liquid–liquid phase separation (LLPS) and interfacial properties of two LLPS modes: associative (ALLPS) and segregative (SLLPS). Analytical expressions for the critical point (CP) and binodal boundaries are derived and show excellent agreement with self-consistent field (SCF) lattice computations.

Distinct thermodynamic features differentiate ALLPS from SLLPS: (1) in ALLPS, polymers co-concentrate within a single dense phase coexisting with a solvent-rich phase, whereas in SLLPS each polymer forms a separate phase; (2) the attractive interaction per monomer in ALLPS is strongly dependent on solvent quality, but solvent-independent in SLLPS; and (3) ALLPS binodals exhibit near-universal behavior, largely independent of solvent content.

SCF results further show that interfacial tension increases and interfacial width decreases with distance from the CP. We provide scaling relations for both quantities are provided. Compared with SLLPS, ALLPS displays higher interfacial tension and a thinner interface, reflecting distinct molecular organization at the liquid–liquid boundary.

\end{abstract}
\clearpage

\noindent

\section{Introduction}
Liquid–liquid phase separation (LLPS) is the demixing of a homogeneous liquid into two (or more) coexisting liquid phases. LLPS occurs in a wide range of colloidal, polymeric, and/or amphiphilic soft matter systems \cite{Anderson2002,Hamley2003,Kwon2023}. Adding a certain amount of nonadsorbing polymers to a colloidal dispersion leads to LLPS \cite{DeHek1979,Gast1986,Ilett1995} if the depletion attraction {\cite{Asakura1954,Binder2014}} is sufficiently long-ranged \cite{lekkerkerker_tuinier_vis_2024}. Self-assembly of 
amphiphiles \cite{Ianiro2019LiquidliquidPS} can be induced by LLPS. Macromolecular LLPS plays a central role in structuring both synthetic \cite{Chengbook} and biological \cite{Wang2021LiquidliquidPS,Monterroso2024} soft matter systems and is used in technological applications \cite{Robeson2010,XuLLPS2021}. In materials science, LLPS involving polymers offers a powerful method to structure matter at the micro- and nanoscale \cite{Franeker2015,XU2023,HuLLPSreview2024}. Through careful design of polymer architecture and interactions, LLPS can generate the spontaneous formation of dynamic, multi-phase materials \cite{Ianiro2019LiquidliquidPS,OBAYASHI2023,Fu2024SupramolecularPF}. 
Importantly, LLPS can be responsive \cite{Xu2021} to external stimuli \cite{Kim2021}—such as temperature, pH, or ionic strength, allowing the creation of adaptive, programmable materials \cite{WANG2021Curr}.

In biology, LLPS has been recognized as a fundamental organizing mechanism in living cells \cite{Minton2000,Juelicher2014,Monterroso2024}. The high concentration of biomacromolecules in cells leads to macromolecular crowding \cite{Walter1995}, which underlies the formation of membraneless organelles (MLOs) \cite{mao_phase_2019,Andre2020} such as nucleoli, stress granules, and other condensates composed of proteins, RNA, and other biopolymers. Biocondensates arising from LLPS \cite{Iborra2007,Uversky2017,Cubuk2022} allow cells to compartmentalize biochemical reactions without lipid membranes, thereby enabling rapid responses to environmental cues. It is increasingly clear that many cellular phase transitions occur close to saturation concentrations, making them easily tunable yet vulnerable to dysregulation. The dense protein phases that arise from LLPS can enhance protein aggregation, which can be regarded as uncontrolled self-organization. Indeed, LLPS has been associated with neurodegenerative diseases \cite{ELBAUMGARFINKLE2019,HAIDER2023} such as amyotrophic lateral sclerosis, Alzheimer's, and Parkinson's, where pathological condensates serve as precursors to toxic protein aggregates \cite{Poudyal2023}. 

While LLPS has attracted considerable attention in recent years, its fundamental thermodynamic origins remain less thoroughly explored, particularly for the associative case. For long chains, binary polymer blends typically phase separate \cite{binder_phase_1994,RubinsteinColby2003,Robeson2014}. When different polymers are mixed in the same solvent, LLPS occurs above a certain polymer concentration \cite{Degennes1979,Wang2019}. In a seminal paper, Dobry and Boyer-Kawenoki \cite{Dobry1947} showed that classical synthetic polymers in a ternary polymer 1 + polymer 2 + solvent 3 mixture are typically incompatible. This can be explained \cite{Vrij1976} by the fact that the entropy of mixing polymer chains is small compared to that of mixing low molar mass substances; however, the mixing enthalpy of most unlike monomers of polymers is positive. Hence, a slight repulsive interaction between different polymers already leads to segregative demixing.

\begin{figure}[htb!]
    \centering
    \includegraphics[angle=0,width=0.67\linewidth]{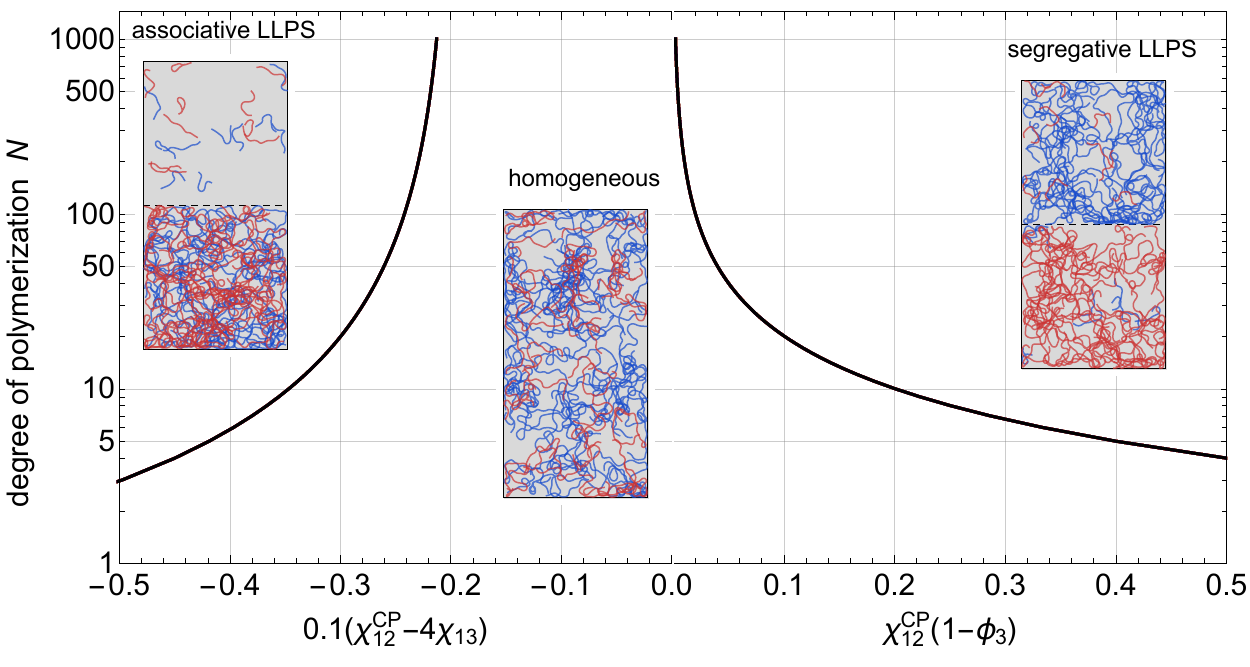}
    \caption{From associative (left) to homogeneous (middle) to segregative (right) Liquid--Liquid Phase Separation (LLPS) state diagram of two symmetric polymers $1$ (blue) and $2$ (red) with chain lengths $N$ in a common solvent (gray). In the segregative case ($\chi_{12} N (1-\phi_3) \geq 2$), one phase (say, $\alpha$) is polymer $1$-rich, whereas the other phase ($\beta$) is polymer $2$-rich. In the associative scenario ($\chi_{12} \leq  4 \chi_{13}-2(1+1/\sqrt{N})^2$), one phase is solvent-rich ($\alpha$) and polymer-lean and both associated polymers concentrate in the other phase ($\beta$). The phase boundaries are indicated in terms of the (scaled) critical polymer--polymer interaction $\chi_{12}^\text{CP}$.} 
    \label{fig:LLPSSketch}
\end{figure}

The opposite case, in which there are (effective) attractions between the different polymers, was investigated at an early stage by Bungenberg de Jong and Kruyt \cite{Bung1929}. It was observed that mixing two solutions containing oppositely charged macromolecules results in associative phase separation, wherein one of the resulting phases is the coacervate phase \cite{Bung1929,Bung1931,Bung1949}. This phase is a dense macromolecular ``complex'' composed of the oppositely charged components. The other phase, the supernatant, is largely devoid of these macromolecular species. Complex coacervation remains widely used to describe such phase behavior \cite{SingPerry2020}, and is, nowadays, for instance, used to prepare artificial cells \cite{Cook2023}.

Although LLPS is relevant across a broad range of disciplines, a clear distinction \cite{Bergfeldt1996,Kruif2001,Crowe2018} between its two principal driving modes is still not consistently made (Fig.~\ref{fig:LLPSSketch}). Segregative liquid-liquid phase separation (SLLPS) is driven by repulsive interactions between dissimilar macromolecules, while associative liquid-liquid phase separation (ALLPS) is governed by attractive interactions between the different polymeric species. This distinction is essential for interpreting the diversity of phase behaviors observed in both synthetic systems and living cells. For example, crowding-induced demixing driven by excluded volume interactions in the bacterial cytoplasm \cite{Walter1995,Uversky2017} exemplifies SLLPS, whereas specific attractive interactions between intrinsically disordered proteins and RNA in eukaryotic organelles \cite{Holehouse2023TheMB} often underlie ALLPS.

In recent years, significant progress has been made to elucidate phase diagrams \cite{Qian2022,Arter2022,Erkamp2023}, enabling the efficient screening of the phase stability of mixtures of different natures. To make progress in interpreting phase diagrams, it is crucial that the experimental toolbox is combined with the theoretical interpretation of phase behavior. 

In this paper, we reveal the differences in phase behavior and related interfacial properties between the two types of LLPS in a unified theoretical mean-field framework. We employ Flory–Huggins theory \cite{Huggins1942,Flory1942} and self-consistent field (SCF) computations \cite{Fleer1998} to study LLPS of binary polymer mixtures in a common solvent,  contrasting SLLPS \cite{binder_phase_1994} with ALLPS. 
For a symmetric (same chain length) binary polymer mixture in a common (same solvency) solvent, we explore SLLPS, ALLPS, and the stable regimes of the mixtures by varying the interaction strength between the two polymers (Fig.~\ref{fig:LLPSSketch}). For this symmetric chain length and (non-selective) solvency case, the advantage is that simple analytical expressions can be derived for the critical points. The theoretical results presented here can be extended to asymmetric polymer mixtures and selective solvents. 

We compute phase diagrams and interfacial properties and demonstrate that these two modes of LLPS exhibit profoundly different phase coexistences. We show that, even though different in magnitude, the scaling relations with distance to the critical of the interfacial tension and width is the same in both modes. Notably, we find that interfacial tension is dramatically higher in the associative case due to the sharp interface between the coacervate and the solvent. We shed light on the role of \textit{effective} solvency $\Delta \chi$: the difference between polymer--polymer and polymer--solvent interaction, on the phase stability region. Our results provide a clear distinction between associative and segregative LLPS, offering insights that are relevant for the design of structured materials and the understanding of intracellular phase behavior. 

\section{Theory}
For a mixture composed of $k$-components $\Delta F^\text{mix}$, the Flory-Huggins (FH) free energy of mixing reads \cite{SouzaStone2024}: $\Delta F^\text{mix}/L k_\mathrm{B}T = f = \sum\limits_{i=1}^k ({\phi_i}/{N_i})\ln{\phi_i}+ \frac{1}{2}\sum\limits_{j \neq i}^k \sum\limits_{i=1}^k {\chi_{ij}\phi_i \phi_j}$, with $\chi_{ij}=\chi_{ji}$ the interaction parameter between components $i$ and $j$, $\phi_i$ the volume fraction of component $i$, $L$ the total number of lattice sites, and $N_i$ the chain length of component $i$. Hence, for a ternary mixture of polymers 1 and polymers 2 with equal chain length $N=N_1=N_2$ in a solvent 3 ($N_{3}=1$), the dimensionless free energy density $f$ per lattice site can be written as \cite{scott_thermodynamics_1949}: 
\begin{equation}
f =  {\frac{\phi_1}{N}\ln{\phi_1} +\frac{\phi_2}{N}\ln{\phi_2}+{\phi_3}\ln{\phi_3}}+ {\chi_{12}\phi_1\phi_2} + {\chi_{13}\phi_3(\phi_1+\phi_2}) ,
\label{FH3comp2}
\end{equation}
where we assumed the solvent quality for both polymers in the solvent is the same: $\chi_{13}=\chi_{23}$, a situation that may be approximated, for instance, for polysaccharides mixed with PEO in water at room temperature \cite{Edelman2003}. Note that the volume filling constraint $\sum\limits_{i=1}^k \phi_i = 1$ is imposed. Here we do not account for the effects of polymer polydispersity on the phase stability of polymer mixtures in a common solvent  \cite{Leuken2023,SouzaStone2024}, we exclude poor solvency cases \cite{Ameslon2025} do not account for solvent disparity \cite{Leermakers2025}. We note that we find that ALLPS follows from FH theory without evoking further attractions between the monomers of the two different polymers \cite{prusty_thermodynamics_2018} or by modifying the entropic terms\cite{dobrynin_phase_2004}.

Analytical expressions for the critical point (CP) of $\chi_{12}$ follow as (see SI): 
\begin{subequations}
  \label{CPFH3chi}
    \begin{empheq}[left={\chi_{12}^\text{CP} =\empheqlbrace\,}]{align}
      & \frac{2}{N (1-\phi_3)}  & \text{SLLPS}
        \label{CPFH3chia} \\
      & 4\chi_{13}-{2\left(1+\frac{1}{\sqrt{N}}\right)^2}  & \text{ALLPS}
        \label{CPFH3chib}
    \end{empheq}
\end{subequations}
The curves in Fig.~\ref{fig:LLPSSketch} follow these expressions. For the associative case, we find that (Eq.~(\ref{CPFH3chib}))  $\chi_{12}^\text{CP,assoc} \leq -2 + 4\chi_{13}$ is required to induce LLPS in the limit $N \rightarrow \infty$. Clearly, $\chi_{12}^\text{CP,assoc}$ depends strongly on solvency $\chi_{13}$ (=$\chi_{23}$) and weakly on chain length $N$, but is independent of $\phi_3$. In a good solvent, it follows that a strong attraction between the two polymers is required ($\chi_{12}^\text{CP,assoc} \leq -2$) to induce a coacervate (a phase of the two dense polymers) coexisting with a solvent-rich phase, whereas for a $\theta$-solvent ($\chi_{13} = 1/2$) a weak attraction suffices to induce demixing ($\chi_{12}^\text{CP,assoc} \leq 0$). This opens up the possibility of ALLPS at relatively small, negative $\chi_{12}$ for long polymer chains in a (close to) $\theta$-solvent, which can be achieved experimentally \cite{Frezzotti1994}.

For the segregative case \cite{
binder_phase_1994}, the classical binary symmetric polymer melt result $\chi_{12}^\text{CP,segr} = 2/N$ (Eq.~(\ref{CPFH3chia})) is recovered \cite{Degennes1979} in the absence of solvent. In the presence of solvent, the specific solvent quality $\chi_{13}$ (=$\chi_{23}$) does not affect the critical point, but the solvent concentration weakens the segregation for entropic reasons \cite{scott_thermodynamics_1949,binder_phase_1994,tromp_polyelectrolytes_2016,Vis2018JPCB,Leermakers2025}.

Analytic expressions for the polymer volume fractions at the CP are:
\begin{subequations}
  \label{CPFH3fi2}
    \begin{empheq}[left={\phi_{1,2}^\text{CP} =\empheqlbrace\,}]{align}
      & \frac{1-\phi_ 3}{2}  & \text{SLLPS}
        \label{CPFH3fi2a} \\
      & \frac{1}{2 \sqrt{N}+2} & \text{ALLPS}
        \label{CPFH3fi2b}
    \end{empheq}
\end{subequations}
Eq.~(\ref{CPFH3fi2a}) reveals that the polymer segment volume fraction at the critical point is independent of chain length for the segregative case. In contrast, Eq.~(\ref{CPFH3fi2b}) indicates that for ALLPS the critical volume fraction is \textit{independent} of the amount of solvent but \textit{depends only} on the polymer chain length. This stems from the fact that the two polymers concentrate in the coacervate phase, coexisting with a nearly pure solvent phase. It is already clear from the critical point expressions above that ALLPS and SLLPS are fundamentally different. This is confirmed by the LLPS phase diagrams. 

\section{Results and Discussion}
By using standard thermodynamics (see SI), we computed the chemical potentials of all components, from which we can compute binodals as a function of $\chi_{12}$, at fixed solvent conditions (via $\chi_{13}$) and solvent concentration $\phi_3$. For a ternary FH mixture, such endeavors have been done before \cite{AltenaSmolders1982,Huang1995,Ameslon2025}, by using only positive Flory-Huggins parameters (SLLPS). For polymers in a binary solvent mixture, Zhang \cite{Zhang2024,Zhang2025} gained useful insights into the co-solvency \cite{Zhang2025} and co-non-solvency \cite{Zhang2024} effects observed experimentally. We focus on two-phase coexistence, while it is recognized that three-phase coexistence may also occur occasionally for poor solvency \cite{Ameslon2025,Leermakers2025}. 

\begin{figure}[htb]
    \centering
    \includegraphics[width=0.77\linewidth]{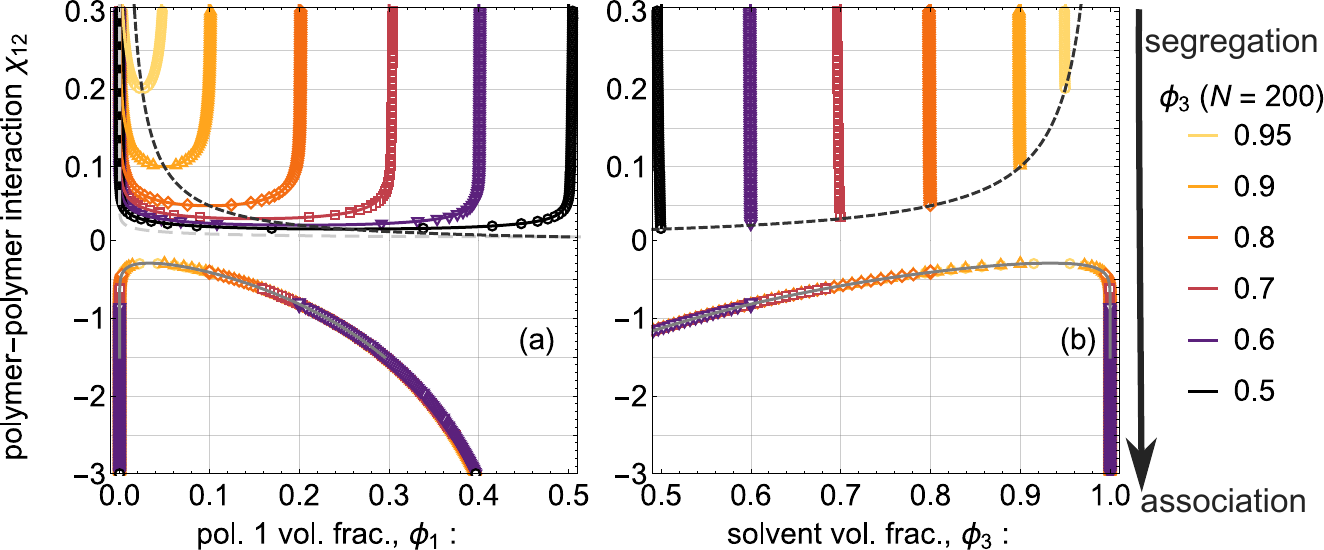}
    \caption{LLPS binodals in the $\chi_{12}$-$\phi_i$ phase space for $N_1=N_2=N=200$ and $N_3=1$ in a $\theta-$solvent ($\chi_{13}=\chi_{23}=0.5$): (a) polymer 1 coexistence volume fractions, $\phi_1$ (from which $\phi_2$ directly follows), (b) solvent coexistence volume fractions, $\phi_3$. Solid curves are calculated FH-binodals, and symbols correspond to SCF-computed binodals. Solvent volume fractions ($\phi_3$) are indicated in the (common) legend. Gray dashed curve correspond to the polymer blend case ($\phi_3 \rightarrow 0$). Black dashed curves correspond to the $\phi_3$-dependent critical points. 
    }
    \label{fig:SCFTheoComparison_N200}
\end{figure}

Fig.~\ref{fig:SCFTheoComparison_N200} shows the phase equilibria in terms of coexisting volume fractions (binodals) for $N=200$ in a $\theta$-solvent ($\chi_{13} = \chi_{23} = 1/2$) for various interaction strengths $\chi_{12}$ between the different polymers. The curves show the phase coexistences calculated from the FH free energy (Eq.~(\ref{FH3comp2})). Data points are the results of numerical SCF lattice computations \cite{Fleer1998,tromp_polyelectrolytes_2016,Vis2018JPCB} that match the FH predictions. SLLPS occurs for sufficiently positive values of $\chi_{12}$. For SLLPS the volume fractions in the coexisting phases are linked through:
\begin{align}
    \phi_1^{\alpha \text{,} \beta} &= 1 - \phi_3 - \phi_2^{\alpha \text{,} \beta} 
    \quad\quad\quad \text{ ; }
    &
    \phi_3^\alpha = \phi_3^\beta \text{ ,} 
\end{align}
\textcolor{black}{where $\alpha$ and $\beta$ are the two coexisting phases considered (see the caption of Fig.~\ref{fig:LLPSSketch})}. Adding more solvent narrows the partitioning of the polymers over the two phases (Fig.~\ref{fig:SCFTheoComparison_N200}(a)) and moves the phase coexistence concentrations to higher $\chi_{12}$ values. The solvent compositions in (Fig.~\ref{fig:SCFTheoComparison_N200}(b)) also follow from (a) through the filling constraint $\phi_1 +\phi_2 +\phi_3 = 1$. We note that the specific solvent quality $\chi_{13}$ (=$\chi_{23}$) does not influence the overall shape of the SLLPS binodals (see SI). For ALLPS, the results in Fig.~\ref{fig:SCFTheoComparison_N200} show that the phase diagram is independent of $\phi_3$. If $\alpha$ is the polymer-rich phase, the relations between component volume fractions for ALLPS are:
\begin{align}
    \phi_{1}^{\alpha\text{,}\beta} &= \phi_2^{\alpha\text{,}\beta}
    \quad\quad\quad \text{ ; }
    & \quad \quad \quad \phi_3^\alpha < \phi_3^\beta . 
\end{align}

\begin{figure}[htb!]
    \centering
    \includegraphics[width=0.7\linewidth]{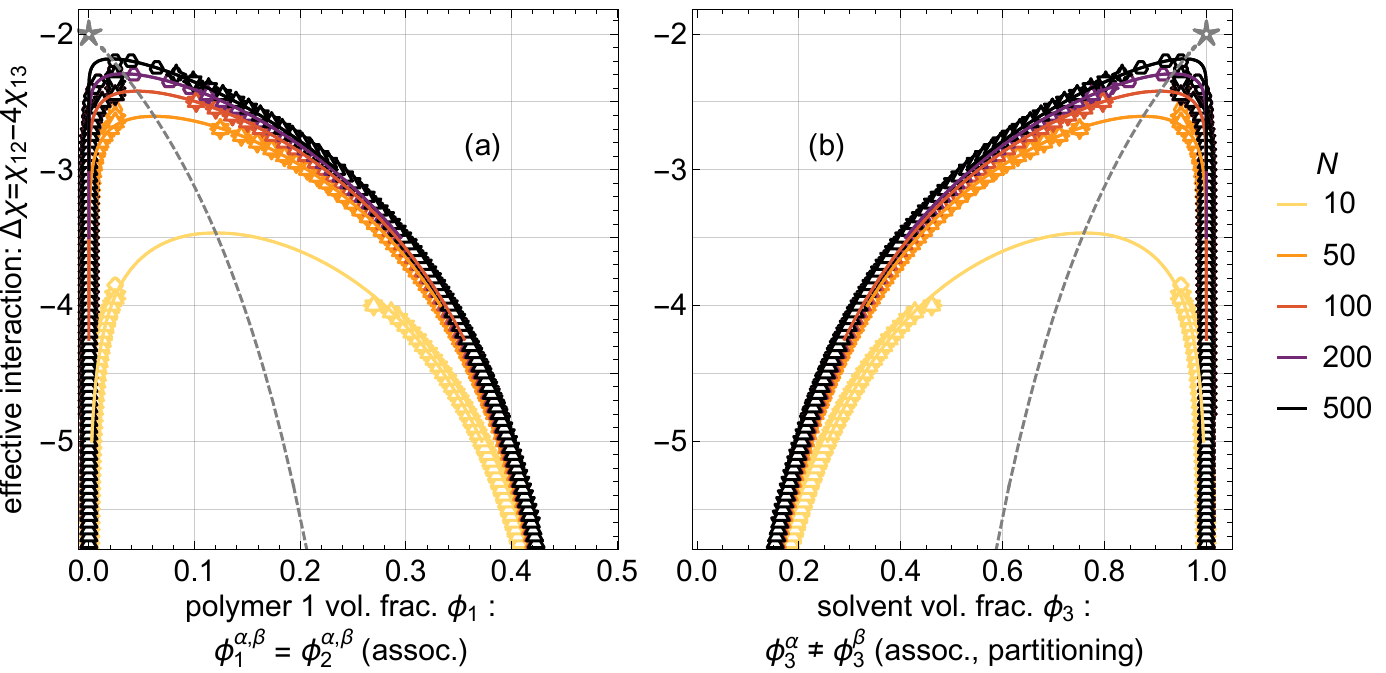}
    \caption{
    ALLPS phase coexistence curves in terms of the effective interaction $\Delta \chi= \chi_{12}-4\chi_{13}$ versus $\phi_i$ for various polymer chain lengths $N$ and solvencies. Solid curves are the theoretical FH binodals. Solid gray dashed curves are the $N$-dependent critical points. SCF results are indicated with colored symbols, with a set of solvency conditions $\chi_{13} = \{0, 0.1, 0.2, 0.3, 0.4, 0.5\}$. The three-pointed gray star symbol represents the critical endpoint in the long-chain limit of the association binodal critical points in common solvent conditions. Different symbols are used for different solvency conditions: $\chi_{13} = 0.5$ (up triangles), $\chi_{13} = 0.4$ (down triangles), $\chi_{13} = 0.3$ (diamonds), $\chi_{13} = 0.2$ (four-pointed stars), $\chi_{13} = 0.1$ (up triangle truncated), $\chi_{13}=0$ (disks). }
    \label{fig:SCFTheoAssocMerged}
\end{figure}

Results for several chain lengths of the ALLPS binodals are presented in Fig.~\ref{fig:SCFTheoAssocMerged} for a range of different solvencies $\chi_{13}$ (different types of symbols). Compared to Fig.~\ref{fig:SCFTheoComparison_N200}, the ordinate is now rescaled: $\chi_{12} \rightarrow (\chi_{12}-4\chi_{13}) \text{ .}$ The effective interaction $\Delta \chi = \chi_{12}-4\chi_{13}$ (which follows from Eq.~(\ref{CPFH3chi})) measures the effective interaction; the direct interaction between the different polymers corrected for the affinity for the solvent. For a given chain length, the binodals for different solvencies collapse onto a single curve.  Fig.~\ref{fig:SCFTheoAssocMerged} emphasizes the fact that the effective interaction, $\Delta\chi$, determines the location of the ALLPS binodals. The segregation binodals are independent of the polymer solvency conditions, as reflected in the independence of their critical point with $\chi_{13}$. Eq.~(\ref{CPFH3fi2a}) explains the volume fraction dependence of the CP (black dashed curves) in Fig.~\ref{fig:SCFTheoComparison_N200}, and Eq.~(\ref{CPFH3fi2b}) gives the chain length dependence (dashed) of the CP in Fig.~\ref{fig:SCFTheoAssocMerged}. The gray stars in Fig.~\ref{fig:SCFTheoAssocMerged} correspond to the critical endpoints at $N \rightarrow \infty$. 

While ALLPS involving complex coacervation is often observed in solutions with oppositely charged polymers \cite{Dubin2011,VEIS2011,Aumiller2016PhosphorylationmediatedRC,SingPerry2020}, ALLPS is commonly not observed for uncharged polymers. Although there are exceptions \cite{Frezzotti1994}, in most cases, the effective interactions between neutral polymers are repulsive, and a slight repulsion is sufficient to induce phase separation above a certain polymer concentration.

In the field of supramolecular chemistry, modern polymer synthesis revealed, however, possibilities to induce strong attractions \cite{Brunsveld2001,Palmans2007} between polymer segments, involving, e.g., hydrogen bonds \cite{Johnson2010} between different monomers. For instance, we mention synthetic nucleobase-containing polymers \cite{Spijker2006,ChengHuangetal2008} that mimic both the structures and functions of natural nucleic acids \cite{Yangnucleo2017}. High-affinity versions of hydrogen bonding polymers have also been developed \cite{Anderson2013}. Mixtures of such polymers can be made so that hydrogen bonds exist between them. In binary mixtures of deep eutectic solvents \cite{Kollau2018}, $\chi_{12}$ values in the range $-5 \lesssim \chi_{12} \lesssim -2$ can describe the strength of hydrogen bond interactions. Hence, it is expected that mixtures of polymers can be prepared that induce ALLPS without opposite charges, but this has not yet been explored. 

Next, we focus on the interfacial properties of the coexisting phases for ALLPS and SLLPS. When LLPS occurs, an interface appears between the phases \cite{EskerVrij1975IncompatibilityIP,Esker1976IncompatibilityOP,Nose1976TheoryOL,binder_collective_1983}, with an interfacial region over which the density profiles of the different components vary. The interfacial tension, resulting from the inhomogeneous density profiles between coexisting phases, plays a central role in determining the morphology and stability of the resulting phases \cite{Aarts2007}. Understanding interfacial properties, such as interfacial tension and thickness, helps in understanding, for instance, the behavior of MLOs, liquid-like compartments within living cells that mediate a range of essential physiological functions. In biological systems, the value of the interfacial tension governs condensate localization via wetting, drives regulated coalescence, and supports the emergence of multiphase organization \cite{Aarts2007,Holland2023} Elucidating how the interactions among the constituent molecules modulate the interfacial tension is, consequently, key in controlling and predicting the behavior of such systems. 

\begin{figure}[htb!]
    \centering
    \includegraphics[width=0.85\linewidth]{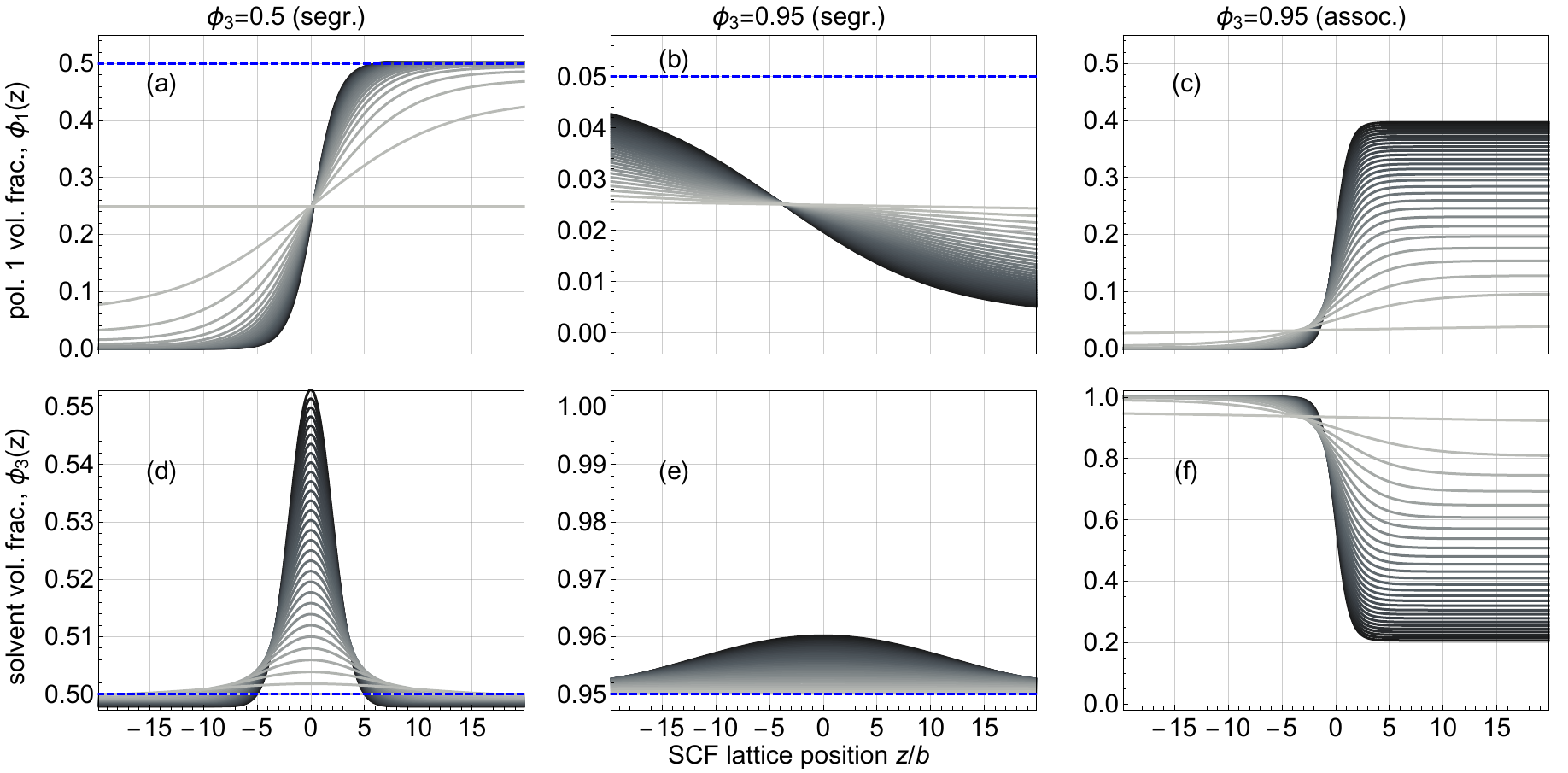}
    \caption{Polymer segment (a-c) and solvent (d-f) density profiles across the interface of demixed polymer 1 - polymer 2 - solvent mixtures for SLLPS (a,b,d,e) and ALLPS (c,f). Curves are fits to SCF computations using Eqs.~(\ref{tanh}) and (\ref{solventads}). Examples correspond to selected phase diagrams in Fig.~\ref{fig:SCFTheoComparison_N200} ($N=200$), so for the $\theta$-solvent condition $\chi_{13}=\chi_{23}=0.5$. In panels (d-f) the solvent concentration is plotted for SLLPS. In the segregative scenario (a,b,d,e), the polymer--polymer interaction parameter $\chi_{12}$ decreases from $0.3$ to the corresponding critical value: for $\phi_3 = 0.5$ (segr.), $\chi_{12} = \{0.3, 0.296,...,0.032,0.028\}$ for $\phi_3 = 0.95$ (segr.), $\chi_{12} = \{0.3, 0.297,...,0.204,0.201\}$ ; (black to light gray as approaching the CP). Blue dashed lines correspond to $1-\phi_3$. In the association case (c,f) $\phi_3 = 0.95$ (assoc.), $\chi_{12} = \{-3.0,-2.9,...,-0.3\}$ (approaching the CP from black to light gray). A systematic comparison of these fitted profiles with the SCF-generated profiles is provided in the SI. 
    }
    \label{fig:SCFprofs1}
\end{figure}

In Fig.~\ref{fig:SCFprofs1}, local polymer segment and solvent volume fraction profiles in the interfacial region are plotted as a function of the position $z$. Near the interface between the coexisting phases $\alpha$ and $\beta$, these curves follow \cite{binder_collective_1983,kerle_effects_1999}:
\begin{equation}
    \phi_i(z) = \frac{\phi_i^{\alpha}-\phi_i^{\beta}}{2} + \frac{(\phi_i^{\alpha}+\phi_i^{\beta})}{2} \tanh{\left(\frac{z-z_0}{w_0}\right)}  \text{ ,}  \label{tanh}
\end{equation}
with $i = \{1,2\}$, $z_0$ the position of the interface and $w_0$, the interfacial width. The numerical SCF lattice results were fitted to Eq.~(\ref{tanh}), giving close agreement (see SI). At fixed solvent concentrations, the collection of profiles varies from light gray to black as the distance to the critical point increases (see the caption for details). 

In the segregative case, we plot the results for solvent volume fractions of 0.5 (a),(d) and 0.95 (b),(e). The interface is sharper when the polymer content is higher. Further from the critical point, where the concentration profiles become increasingly inhomogeneous, the interfacial width decreases. 

Remarkable is the local adsorption of solvent at the interface for SLLPS (Fig.~\ref{fig:SCFprofs1}(d),(e)). This positive adsorption can be explained \cite{Vis2018JPCB} by the fact that the solvent is not repelled by either of the polymers; however, the polymers do repel each other. Therefore, the solvent can reduce the number of unfavorable polymer 1 -- polymer 2 contacts by accumulating at the interface. 
This $z$-dependent adsorption profile near the interface at $z=z_0$ can be described using a shifted normal distribution:
\begin{equation}
    \phi_3 (z) = \phi_3^\text{ads}  \mathrm{e}^{-{\frac{(z-z_0)^2}{4w_0}}} +\frac{\phi_3^\alpha + \phi_3^\beta}{2} \text{ ,}
\label{solventads}
\end{equation}
where $\phi_3^\text{ads}$ quantifies the amplitude of the Gaussian function and relates to the adsorption of the solvent. The solvent concentrations in the bulk phases $\alpha$ and $\beta$ are given by $\phi_3^\alpha$ and $\phi_3^\beta$. It is known that the combination of two hyperbolic functions leads to a normal distribution \cite{tsai_hyperbolic_2017}. In the SI, we show that Eq.~(\ref{solventads}) accurately describes SCF data. 

In the associative case (Fig.~\ref{fig:SCFprofs1}(c),(f)) the interface looks fundamentally different. The interface is relatively sharp, also far from the critical point, even at this low polymer content ($\phi_3=0.95$). For SLLPS (b,e) ($\phi_3=0.95$) the interfacial region, where the concentration profiles are inhomogeneous, is relatively wide (spanning the entire range shown here), whereas it is much thinner for ALLPS (c,f). In the associative case, one phase mainly contains solvent, while the other is dense in both polymers. Conversely, in the segregative case, both phases contain (for the symmetric case equal and) significant amounts of polymers that repel each other, which explains the wider interface. Note that the observed adsorption peaks are consistently less pronounced for $\chi_{13}=0$  (see SI) compared to $\Theta$-solvent conditions. 

All ALLPS polymer segment concentration profiles (Fig.~\ref{fig:SCFprofs1}) follow Eq.~(\ref{tanh}), with $\phi_1(z)\approx \phi_2(z)$: the two polymers associating in a single solvent-lean phase, and the solvent partitioning into a solvent-lean (polymer-rich) and a solvent-rich (polymer-lean) phase. As observed in (f), the solvent concentration in the polymer-rich phase decreases the further away $\chi_{12}$ is from the CP. For instance, for the black curve, $\chi_{12}=-3$, $\phi_3^\alpha\approx 0.4$ (the polymer-rich phase far away from the interface, see the corresponding binodal in Fig. \ref{fig:SCFTheoComparison_N200}(b)). 

\begin{figure}[htb!]
 \centering
    \includegraphics[width=0.525\textwidth]{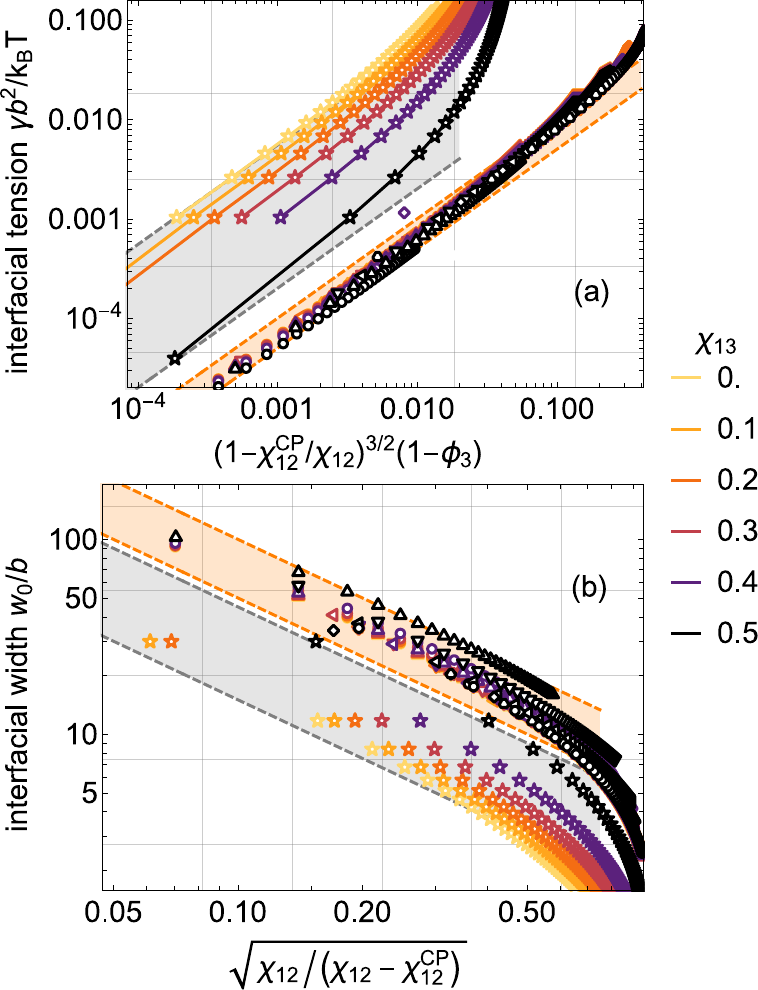}
    \caption{Interfacial tension $\gamma$ (a) and interfacial width $w_0$ (b) for polymer 1 - polymer 2 - solvent mixtures with $N_1=N_2=200$. Numerical SCF lattice computations (data) for both SLLPS (gray symbols) and ALLPS (colored symbols) follow the scaling laws of Eq.~(\ref{gammascaling}) (a) and Eq.~(\ref{width}) (b). Star symbols correspond to the ALLPS interfacial properties; other symbols correspond to SLLPS for various solvent concentrations: $\phi_3=0.95$ (circles), $\phi_3=0.9$ (up triangles), $\phi_3=0.8$ (down triangles), $\phi_3=0.7$ (left triangles)\, $\phi_3=0.6$ (diamonds), and $\phi_3=0.5$ (up truncated triangles). Scaling relations are indicated on the abscissa. The boundaries of scaling ranges for the surface tension (a) in terms of the system parameters are $\gamma = (0.2-5.5) \left(1-{\chi_{12}^\text{CP}}/{\chi_{12}}\right)^{3/2}(1-\phi_3)$ (association, increasing with decreasing $\chi_{13}$), and $\gamma = (0.05-0.1) \left(1-{\chi_{12}^\text{CP}}/{\chi_{12}}\right)^{3/2}(1-\phi_3)$ (segregation, practically independent of $\chi_{13}$). For the interfacial widths (b), the boundaries in terms of the scaling relation are $w_0 = (4.5-1.5)\sqrt{\chi_{12}/(\chi_{12}-\chi_{12}^\text{CP})}$ (association, increasing with increasing $\chi_{13}$) and $w_0 = (5-10)\sqrt{\chi_{12}/(\chi_{12}-\chi_{12}^\text{CP})}$ (segregation).
    }
    \label{fig:InterfScale}
\end{figure}

The numerical SCF lattice computations also yield (see SI) the interfacial tension $\gamma$ of the interface resulting from the LLPS. The SCF fits to Eq.~(\ref{solventads}) provide the interfacial thickness, $w_0$. The polymer--polymer interaction ($\chi_{12}$) and the solvent concentration ($\phi_3$) turn out to be the two main tuning knobs on the LLPS landscape determining the interfacial properties. We elucidated the dependencies of $\gamma$ and $w_0$ on $\chi_{12}$ and $\phi_3$, by making use of previously derived relations for solvent-free binary polymer mixtures \cite{kerle_effects_1999}. Our main findings are summarized in Fig.~\ref{fig:InterfScale}. In the SI (Fig.~S5) we provide the raw SCF data.

Remarkably, the scaling behavior of both the interfacial tension ($\gamma$) and the interfacial width ($w_0$) is independent of the nature of the LLPS type: only their magnitude (the prefactor of the scaling relation) is affected by the nature of the phase separation. As observed already in Fig. \ref{fig:SCFprofs1}, the interface is sharper for ALLPS compared to SLLPS. This explains why we find a much higher interfacial tension for ALLPS (Fig.~\ref{fig:InterfScale}(a)). 

We find that the following scaling relation of the interfacial tension $\gamma$:
\begin{align}
    \gamma \sim \left(1-\frac{\chi_{12}^\text{CP}}{\chi_{12}}\right)^{3/2}(1-\phi_3),\label{gammascaling}
\end{align}
describes the SCF computations. Eq.~(\ref{gammascaling}) is based on the well-known result \cite{kerle_effects_1999,Binder2005}  for solvent-free polymer mixtures, $\gamma \sim (1-\chi_{12}^\text{CP}/\chi_{12})^{3/2}$. Additionally, accounting for the solvent concentration dependence, via $1-\phi_3$, suffices to describe $\gamma$ using Eq.~(\ref{gammascaling}). This $\phi_3$-dependency arises because the interfacial tension is directly related to the polymer concentration difference across the interface; consequently, increasing the solvent content leads to a reduction in interfacial tension for both LLPS modes. For SLLPS (orange regions), the interfacial tension appears to be independent of the specific common polymer solvent solubility conditions $\chi_{13}$ and of the solvent concentration $\phi_3$ with varying polymer--polymer interaction $\chi_{12}$ when using the scaling of Eq.~(\ref{gammascaling}), where the critical point plays a crucial role. For ALLPS  (gray regions), there is a decrease in the interfacial tension as $\chi_{13}$ increases from 0 to 0.5.

For $w_0$, we expect \cite{kerle_effects_1999}
\begin{align}
    w_0 \sim \sqrt{\frac{{\chi_{12}-\chi_{12}^\text{CP}}}{{\chi_{12}}}} \text{ ,}
    \label{width} 
\end{align}
and we find that this holds quite generally for a wide range of (common) solvency conditions for two polymers, accounting for both SLLPS and ALLPS (Fig.~\ref{fig:InterfScale}(b)). There is a weak polymer solvency dependence of the interfacial width $w_0$ for both SLLPS and ALLPS. The small values for $\gamma$ correspond to large $w_0$ values.

We note that the Flory-Huggins theory used in this work, which is the basis of SCF, is crude. The enthalpy expression used neglects correlation effects in the occupancy of lattice sites and 
overestimates the number of neighbors of polymer segments in the lattice. In the FH approach, it is assumed that the lattice sites can be occupied independently of each other. This is not correct for chains that must be both self-avoiding and mutually avoiding, as it neglects the disparity in size and shape of the subunits of the two types of chains in a polymer blend, along with packing constraints and specific interactions. Even within the realm of a lattice model, the statistical mechanics involve rough approximations beyond the mean field approximation. As a result, the FH scaling prediction differs from the correct critical behavior. Flory-Huggins theory, however, appears to be qualitatively correct in many circumstances \cite{binder_phase_1994}. It can, therefore, be used to focus on the `universal' aspects of the phase behavior of polymer mixtures, which is exactly the purpose of this work. 

\section{Concluding remarks}
We have shown that one can realize liquid-liquid phase separation (LLPS) of a binary polymer mixture in a common non-selective solvent through two rather different driving forces. Segregative LLPS (SLLPS) occurs if there are any strong repulsive interactions between the components, i.e., if any of the $\chi$'s is sufficiently large, even if the other $\chi$'s are small. Associative LLPS (ALLPS) occurs if the interactions between the components are sufficiently different, i.e., if $\Delta \chi$ of one (or more) of the combinations of interactions is large. This holds true for both favorable and unfavorable interactions; it is driven by $\Delta \chi$. This implies that phase separation is also possible in the case of a mixture with only favorable interactions. The latter insight also explains the findings of Zhang \cite{Zhang2024,Zhang2025} on the co-solvency and co-non-solvency of a polymer solution in a binary liquid mixture using the same Flory-Huggins approach for a ternary mixture. 

The ALLPS phase equilibria are quite different from those for SLLPS because, in ALLPS, both polymers concentrate in a single phase, and the amount of solvent does not affect the equilibrium. SLLPS binodals strongly depend on the amount of solvent, which tunes the partitioning between the two different repulsive polymers. The observation that, unlike SLLPS, ALLPS equilibria are independent of solvent concentration can be used to experimentally determine whether SLLPS or ALLPS has occurred.

We showed that tie-lines do not directly provide information about interactions; rather, they provide information about the component distributions, which, in turn, are a result of the monomer--monomer interactions and solvent concentration. 

We find that the scaling of interfacial tension and width with respect to the distance to the critical point appears to be universal, independent of the mode of LLPS. The nature of phase coexistence and the interface is, however, heavily dependent on the type of LLPS. The ALLPS interface is much sharper compared to SLLPS. leading to significantly higher interfacial tensions of ALLPS than those of SLLPS. The solvent quality tunes the interfacial properties for ALLPS; with better solvent quality, the interfacial tension increases and the interface becomes sharper. 



%
\bibliographystyle{sciencemag}

%
%
%
%
%
%


\section*{Acknowledgments}
We thank drs C. M. (Max) Martens, Nadia A. Erkamp (TU/e), Vangelis Karagianniakis (International School of Utrecht), and professors Jan van Hest, Anja Palmans, Mark Vis (TU/e), Klaus Huber (Paderborn University), and Sissi de Beer (University of Twente) for useful discussions.
\paragraph*{Funding:}
This work was financially supported by the Dutch Ministry of Education, Culture and Science (Gravity Program 024.005.020 – Interactive Polymer Materials IPM).
\paragraph*{Author contributions:}
Conceptualization: R.T. and A.G.G. Methodology: R.T. and A.G.G.. Resources: R.T. Software: A.G.G. Investigation: R.T. and A.G.G. Formal analysis: A.G.G. and R.T. Visualization: R.T. Data curation: A.G.G. Validation: R.T. and A.G.G. Writing—original draft: R.T. and A.G.G. Writing—review and editing: R.T. and A.G.G. Project administration: R.T. Supervision: R.T. Funding acquisition: R.T.

\paragraph*{Competing interests:}
There are no competing interests to declare.
\paragraph*{Data and materials availability:} Wolfram Mathematica was used for the analytical calculations and the data processing. Our Wolfram Mathematica notebooks are available upon request. We used the SFbox provided by F.A.M. Leermakers to compute the SCF results. Our input files and data are available upon request. See also the https://github.com/leermakers/Namics by F. A. M. Leermakers, R. Varadharajan, D. Emmery and A. Kazakov, Namics – MD-SCF Hybrid simulation tool, 2023 for a computer code to perform SCF computations.

\subsection*{Supplementary materials}
Calculation of binodals and critical points from Helmholtz free energy\\
Numerical Self-Consistent Field Lattice Computations\\
Phase diagrams for $\chi_{13}=0$\\
SCF volume fraction profiles\\
(Unscaled) interfacial properties 


\newpage


\renewcommand{\thefigure}{S\arabic{figure}}
\renewcommand{\thetable}{S\arabic{table}}
\renewcommand{\theequation}{S\arabic{equation}}
\renewcommand{\thepage}{S\arabic{page}}
\setcounter{figure}{0}
\setcounter{table}{0}
\setcounter{equation}{0}
\setcounter{page}{1} 


\begin{center}
\section*{Supplementary Materials for\\ \scititle}

Remco~Tuinier$^{1,\ast}$,
{\'A}lvaro~Gonz{\'a}lez~Garc{\'i}a$^{1,2}$\and \\
\small$^\ast$Corresponding author. Email: r.tuinier@tue.nl\and

\end{center}

\subsubsection*{This PDF file includes:}
Calculation of Binodals and Critical Points \\
Numerical Self-Consistent Field Lattice Computations\\
Phase diagrams for $\chi_{13} = 0$\\
SCF volume fraction profiles across the interface\\
(Unscaled) Interfacial properties\\


\newpage

This Supporting Information contains more detailed information on the theoretical methods to calculate phase coexistence using the analytical theory, provides background on the numerical SCF lattice computations to compute binodals and interfacial properties, and includes additional results and further detailed information. 
Mathematica notebooks used for the calculations in this paper are available on request.

\subsection*{Calculation of binodals and critical points from Helmholtz energy for a ternary mixture}
Below we provide the thermodynamic Flory-Huggins (FH) theory expressions 
for a three-component mixture:
\begin{align}
    \frac{F}{k_\mathrm{B} T} = &  n_1 \ln \phi_1 + n_2 \ln \phi_2 + n_3 \ln \phi_3 + \nonumber  \\
    & \chi_{12} \frac{n_1 N_1 n_2 N_2}{n_1 N_1 + n_2 N_2 + n_3 N_3} + \chi_{13} \frac{n_1 N_1 n_3 N_3}{n_1 N_1 + n_2 N_2 + n_3 N_3} +  \label{FHFn}\\ &\chi_{23} \frac{n_1 N_1 n_3 N_3}{n_1 N_1 + n_2 N_2 + n_3 N_3}, \nonumber
\end{align}
where $n_i$ is the number of molecules of component $i$. Following the notation of van Leuken \textit{et al.} \cite{Leuken2023} 
we can write down the FH free energy density $f={F}/{L k_\mathrm{B} T}$ expression (i.e., the free energy per lattice unit) for such a three-component mixture as:
\begin{align}
 f =  {\frac{\phi_1}{N_1}\ln{\phi_1} +\frac{\phi_2}{N_2}\ln{\phi_2}+\frac{\phi_3}{N_3}\ln{\phi_3}}+ {\chi_{12}\phi_1\phi_2} + {\chi_{13}\phi_2\phi_3+\chi_{23}\phi_2\phi_3} \text{ ,}
\label{FH3comp2SI}
\end{align}
with:
\begin{align*}
    \phi_i = \sum_{i=1}^3 \phi_i = 1 \text{ ; }  \phi_i = \frac{n_i N_i}{L} \text{ , with  } L = \sum_{i=1}^3 n_i N_i
\end{align*}
The chemical potentials for the three components follow from:
\begin{align}
    \mu_i = \left(\frac{\partial F}{\partial n_i}\right)_{T, L} \text{ ,}
\end{align}
where $n_i$ is the number of molecules of component $i$ in the system. Implicitly, symmetric contact interactions between different species have been considered, whereas the self-interactions are regarded as athermal (only excluded volume interactions): 
\begin{align*}
    \chi_{ij} = \chi_{ji} \text{ , } \chi_{ii} = 0 
\end{align*}
When taking the derivatives of $F$ with respect to $n_i$ to find $\mu_i$, we found that the most convenient approach is to use Eq.~\eqref{FHFn}. For a ternary mixture, the chemical potentials read:
\begin{align}
    \frac{\mu_i}{k_\mathrm{B} T} = \tilde{\mu_i} = \tilde{\mu_i}^*+\ln\phi_i+(1-\phi_i) - N_i \sum_{j \neq i} \left[\frac{\phi_j}{N_j} - \phi_j(1-\phi_i)\chi_{ij} + \frac{1}{2}\sum_{j\neq k} \phi_j\phi_k \chi_{jk}\right] \text{ ,}
\end{align}
where $\tilde{\mu_i}^*$ is the reference chemical potential. With adequate boundary conditions, binodals can now be computed using the phase coexistence conditions that the chemical potentials for all three components are equal in both phases.

SLLPS calculations were carried out by fixing the \textit{total} concentration ($c$) of the solvent, $c=\phi_3^\alpha+\phi_3^\beta$  \cite{AltenaSmolders1982,binder_phase_1994,Huang1995,Choi2016}. Thus,we have a nonlinear system of 6 equations and 6 unknowns:
\begin{align}
    \tilde{\mu}_i^\alpha = \tilde{\mu}_i^\beta \text{ , }
    \sum_{i=3}\phi_i^\alpha = 1 \text{ , }
    \sum_{i=3}\phi_i^\beta = 1 \text{ , }
    \phi_3^\alpha + \phi_3^\beta = c \text{ .}
\end{align}
which is solved using the {``FindRoot[]''} option in Wolfram Mathematica \cite{MathematicaV12}.  

In the segregative case, a simplified free energy expression can be derived using $\phi_1+\phi_2+\phi_3 = 1$, denoting $\phi_1 = \phi$:
\begin{align*}
    f^\text{segr}= & \frac{\phi\ln\phi}{N}+ \frac{(1-\phi-\phi_3)\ln(1-\phi-\phi_3)}{N} + \phi_3\ln\phi_3 \\
    & + \phi(1-\phi-\phi_3)\chi_{12} + \phi_3(1-\phi_3)\chi_{13}
\end{align*}
The expression above recovers the well-established \cite{Flory1953} 2-component FH expression for a symmetrical polymer blend (in the limit $\phi_3 \rightarrow 0 $):
\begin{align}
    f=\frac{\phi\ln\phi}{N}+\frac{(1-\phi)\ln(1-\phi)}{N} + (1-\phi)\phi\chi_{12} 
\end{align}
for vanishing solvent concentrations, it reflects the full concentration asymmetry of components 1 and 2.

For ALLPS, binodals were calculated directly using the well-known common tangent construction from the derived free energy expressions. In a subsequent follow-up paper, we also intend to present and discuss common tangent plots in detail.

In the associative case, we assume $\phi_1 = \phi_2 = \phi$ in each phase (hence, $\phi_3 = 1-2\phi$), leading to:
\begin{align}\label{eq:f3CassSym}
    f^\text{assoc}= \frac{2\phi \ln\phi}{N}+  (1-2\phi)\ln(1-2\phi) + \chi_{12}\phi^2 + 2(1-2\phi)\phi\chi_{13} \text{ .}
\end{align}
Since there is only one relevant variable in this specific case for both the segregation and associative scenario (one of the polymers' volume fractions, $\phi$), the critical point $(\phi^\text{crit},\chi^\text{crit})$ can be realized by simply looking at the conditions where:
\begin{align}
        \frac{\mathrm{d}^2 f}{\mathrm{d}\phi^2} = \frac{\mathrm{d}^3 f}{\mathrm{d}\phi^3} = 0 .
\end{align}
Using the corresponding free energy expressions, we obtain the expressions presented in the main text for the critical points. All phase equilibria calculations were conducted using Wolfram Mathematica. The scripts used to generate the data are available upon request.

\subsection*{Numerical Self-Consistent Field Lattice Computations}
Self-consistent field (SCF) computations were employed to compute the thermodynamic properties of solutions containing a ternary mixture of polymer 1 + polymer 2 and solvent 3 \cite{Leermakers2025}. Here, we outline the implementation used in this work, which is based on the numerical lattice SCF theory approximation developed by Scheutjens and Fleer \cite{Scheutjens1979,Scheutjens1980,Fleer1998,FleerCohenStuartLeermakersLyklemaFICS5Ch1}.

In the Scheutjens–Fleer self-consistent field (SF–SCF) method, space is discretized into a lattice with coordination number $Z$. The lattice is composed of a set of lattice sites, with each molecular segment occupying a single site. 

For the systems considered here, a single concentration gradient suffices, as the relevant variations occur only along the direction normal to a liquid–liquid interface. To study liquid–liquid phase separation, we adopt a flat geometry. The $z$-coordinate indexes lattice layers, numbered $z=0,1,2,\ldots,L$, with reflecting (mirror) boundaries imposed at $z=0$ and $z=L+1$. The system size $L$ is chosen sufficiently large to ensure that bulk concentrations are reached at both boundaries.

In SCF theory, the Helmholtz energy $F$ is expressed as a functional of the volume fractions $\phi_i(z)$ of each component $i$ and their corresponding local potentials $u_i(z)$ \cite{Fleer1998}:
\begin{align}
\frac{F[{\phi},{u},\alpha]}{k_\mathrm{B}T} = &-\ln Q([{u}]) - \sum_z \sum_i \frac{u_i(z)}{k_\mathrm{B}T} \phi_i(z) + \nonumber \\ &\frac{F^{\rm int}([{\phi}])}{k_\mathrm{B}T} + \sum_z \alpha(z) \left( \sum_i \phi_i(z) - 1 \right),
\label{F}
\end{align}
where $\alpha(z)$ is a Lagrange multiplier enforcing local incompressibility, $\sum_i \phi_i(z) = 1$. The index $i$ labels the different components in the system, which here include solvent molecules and two types of polymer chains. Polymers are modeled as linear chains, with each segment occupying one lattice site of size $b$.

The first term in Eq.~\eqref{F} is the partition function $Q({u},V,T)$. To evaluate it, the molecular architecture must be specified. The total partition function factorizes into single-chain contributions:
\begin{equation}
Q = \prod_i \frac{(q_i)^{n_i}}{n_i!},
\end{equation}
where $q_i$ is the single-molecule partition function for species $i$. The freely jointed chain model is employed, which enables the use of an efficient propagator formalism \cite{Scheutjens1979}. The SCF computations employ the Edwards formalism, where the spatially inhomogeneous equilibrium structures formed by polymer liquids are elucidated by solving the Edwards diffusion equation for polymer chain statistics in position-dependent potential fields.

For a molecule $i$ composed of $N_i$ segments labeled $s=1,\ldots,N_i$, with $\delta_{is}^i = 1$ if segment $s$ is of type $i$ (and zero otherwise), the potential energy of a specific conformation $c$ is
\begin{equation}
\frac{u_i^c}{k_\mathrm{B}T} = \sum_s \sum_i \delta_{is}^i, u_i(r_{is}^c),
\end{equation}
where $r_{is}^c$ is the position of segment $s$ in conformation $c$. The single-chain partition function is then
\begin{equation}
q_i = \sum_c \exp\left(-u_i^c\right).
\end{equation}
Direct evaluation is intractable, as the number of conformations scales as $Z^{N_i-1}$. The propagator formalism circumvents this by reducing the computational cost to scale linearly with $N_i$, while simultaneously yielding segment-level volume fraction profiles $\phi_i(r,s)$. Summing over all segments of type $i$ gives $\phi_i(r) \equiv \phi_i[u]$.

The second term in Eq.~\eqref{F} is a Legendre transformation, converting the system of interest to the $(N,V,T)$ ensemble. The third term, $F^{\rm int}$, contains all inter-segment interactions. Nearest-neighbor interactions are accounted for within the Bragg–Williams mean-field approximation \cite{Flory1953,Hill1962}, parameterized by Flory–Huggins interaction parameters $\chi_{ij}$.

Minimization of $F$ with respect to $\phi_i(z)$ yields
\begin{equation}
u_i(z) = \alpha(z) + \frac{\partial F^{\rm int}}{\partial \phi_i(z)},
\end{equation}
which, together with the propagator formalism, is solved self-consistently. In the mean-field approximation, interactions within a layer are spatially averaged. For example, 
$\langle \phi_i(z) \rangle$ is the nearest-neighbor average of the volume fraction of component $i$:
\begin{equation}
\langle \phi_i(z) \rangle = \lambda_1 \phi_i(z-1) + \lambda_0 \phi_i(z) + \lambda_1 \phi_i(z+1),
\end{equation}
with $\lambda_0 /2 = \lambda_1 = 1/Z$ for a planar lattice. In this work, we took $\lambda_1 = 1/3$. For more details on SCF, see Fleer \textit{et al.} \cite{Fleer1998}.

The grand potential is obtained from the Helmholtz energy by subtracting the chemical work:
\begin{equation}
\Omega = F - \sum_i \mu_i n_i,
\end{equation}
where $\mu_i$ is the chemical potential of species $i$ in the bulk (composition $\phi_i^{\mathrm{b}}$). From $\Omega$, the interfacial tension $\gamma$ between coexisting phases follows directly \cite{tuinier_interfacial_2012,Leermakers2025} from $\Omega$ in a normalized manner as $\gamma b^2/k_\mathrm{B}T$.

In all SCF results shown, the number of lattice sites used is $L = 400$. The initial polymer concentration distributions differ in the segregative and associative scenarios. For the segregative case, polymers of types (1) and (2) are initially pinned near the mirrors of the lattice ($z=0$ and $z=L$). For association, both polymers were initially pinned near the upper boundary of the lattice, $z=L$. The iterative solution provided by SFbox achieves self-consistency between the monomer density profiles and the effective potential fields \cite{Martens2022DeplInt}. Example SFbox input files are available upon request. 

The 
SCF results are compared to 
the Flory-Huggins theory predictions. 
The component concentrations, far away from the interfaces formed on the lattice, are compared with the theoretical FH binodals. 

SCF allows for the calculation of interfacial properties from the equilibrium grand potential values, providing a thermodynamically consistent framework for analyzing phase separation and interfacial phenomena in multicomponent polymer systems \cite{tuinier_interfacial_2012,tromp_polyelectrolytes_2016}.

All data generated from SFbox were analyzed using Wolfram Mathematica. Scripts for Data Analysis are available upon request.

\subsection*{Phase diagrams for $\chi_{13}=0$}
In the main text, we focused on the $\theta$-solvent case $\chi_{13}=0.5$, as it is theoretically an interesting condition, but it is also close to the experimental conditions of soluble polymers (often in the range $0.3 \lesssim \chi_{13} \lesssim 0.6$). Another useful theoretical limit is the athermal, good solvent condition $\chi_{13}=0$.
\begin{figure}[htb!]
    \centering
    \includegraphics[width=0.9\linewidth]{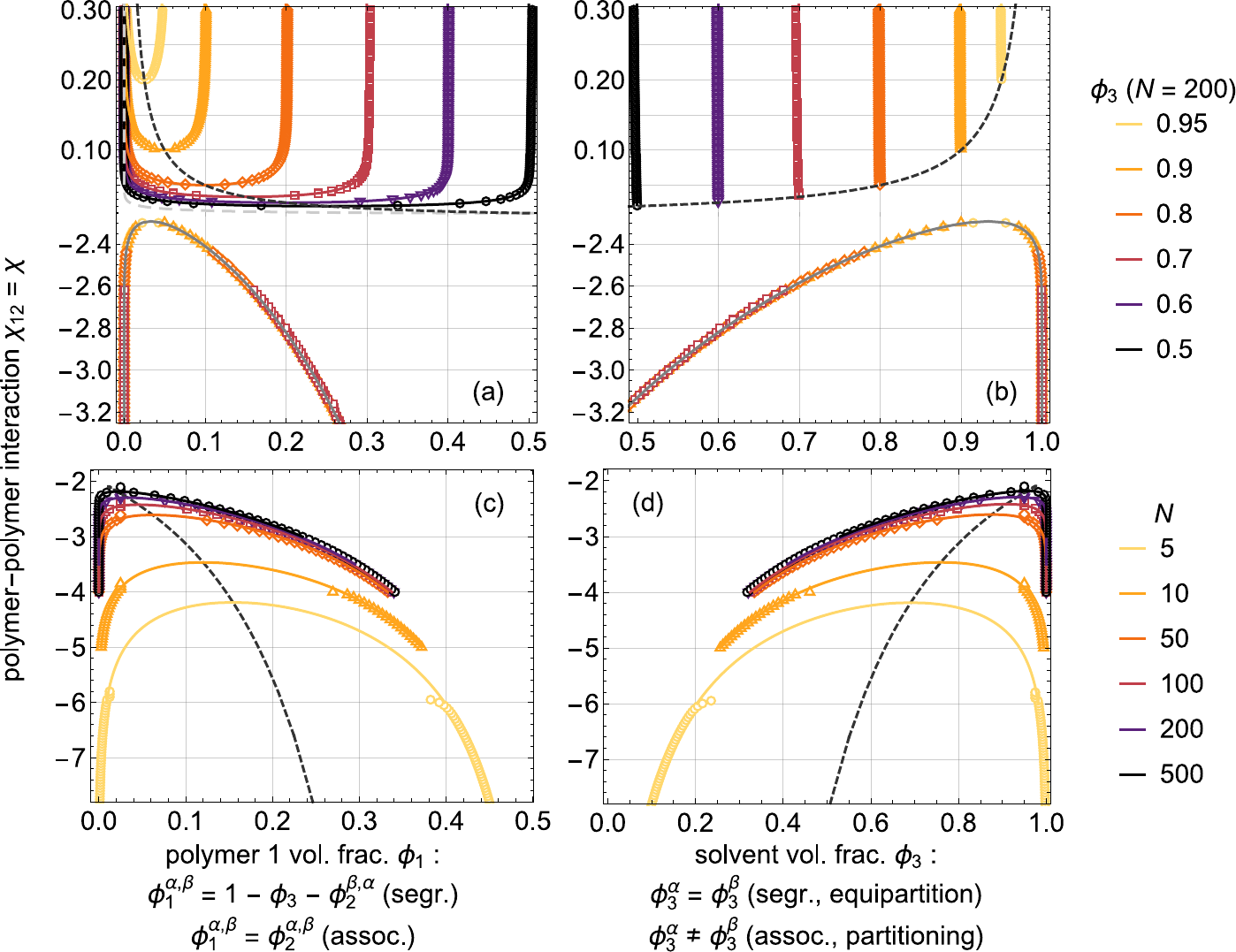}
    \caption{Similar to Figs. 1 and 2, but for $\chi_{13}=\chi_{23}= \chi_{ps} = 0$. Other parameters as Figs. 1 and  2.}
    \label{fig:SCFTheoComparison_N200_ChiPS0}
\end{figure}

\subsection*{SCF volume fraction profiles across the interface}
\begin{figure}[htb!]
    \centering
    \includegraphics[width=0.9\linewidth]{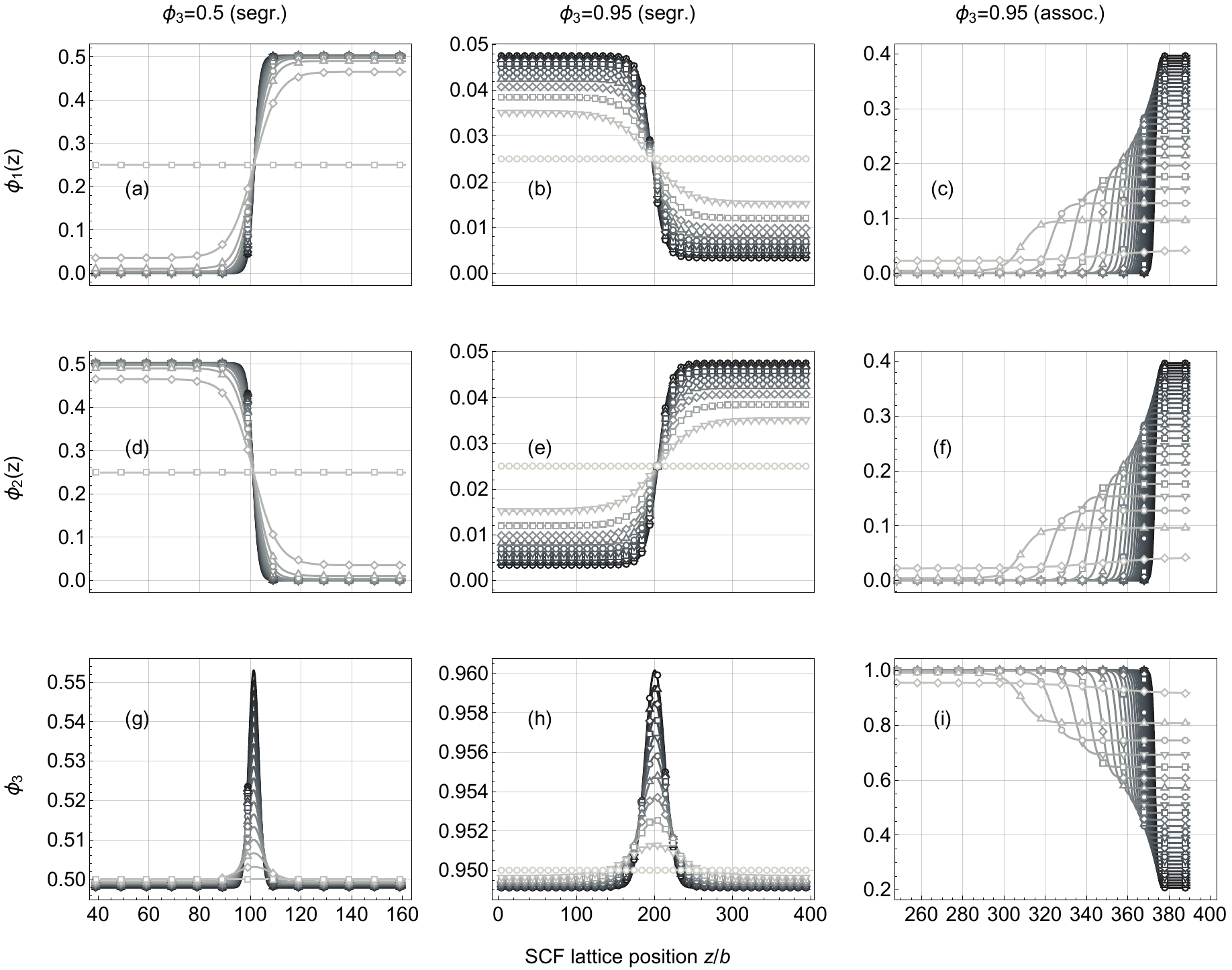}
    \caption{Concentration profiles as provided by SCF with the corresponding fitted functions. Color scale goes from black to light gray approaching the critical point. The specific set of polymer--polymer interaction parameters considered are the same as in Fig. 5. For visualization purposes, a subset of the SCF data (every tenth point) is plotted to prevent the dense data from obscuring the theoretical curves and to more clearly illustrate the agreement between them.}
    \label{fig:SCFProfsFit}
\end{figure}

Fig.~\ref {fig:SCFProfsFit} shows a collection of SCF concentration profiles for the three components in the system, as computed using Scheutjens-Fleer SCF with the SFbox software. Fitting curves are nearly imperceptible as they overlap with SF-SCF data points. The $\chi_{12}$-values considered are the same as those in Fig. 5 in the main text. It is clear that equations (6) and (7) accurately describe the SCF data.

\begin{figure}[htb!]
    \centering
    \includegraphics[width=0.99999\linewidth]{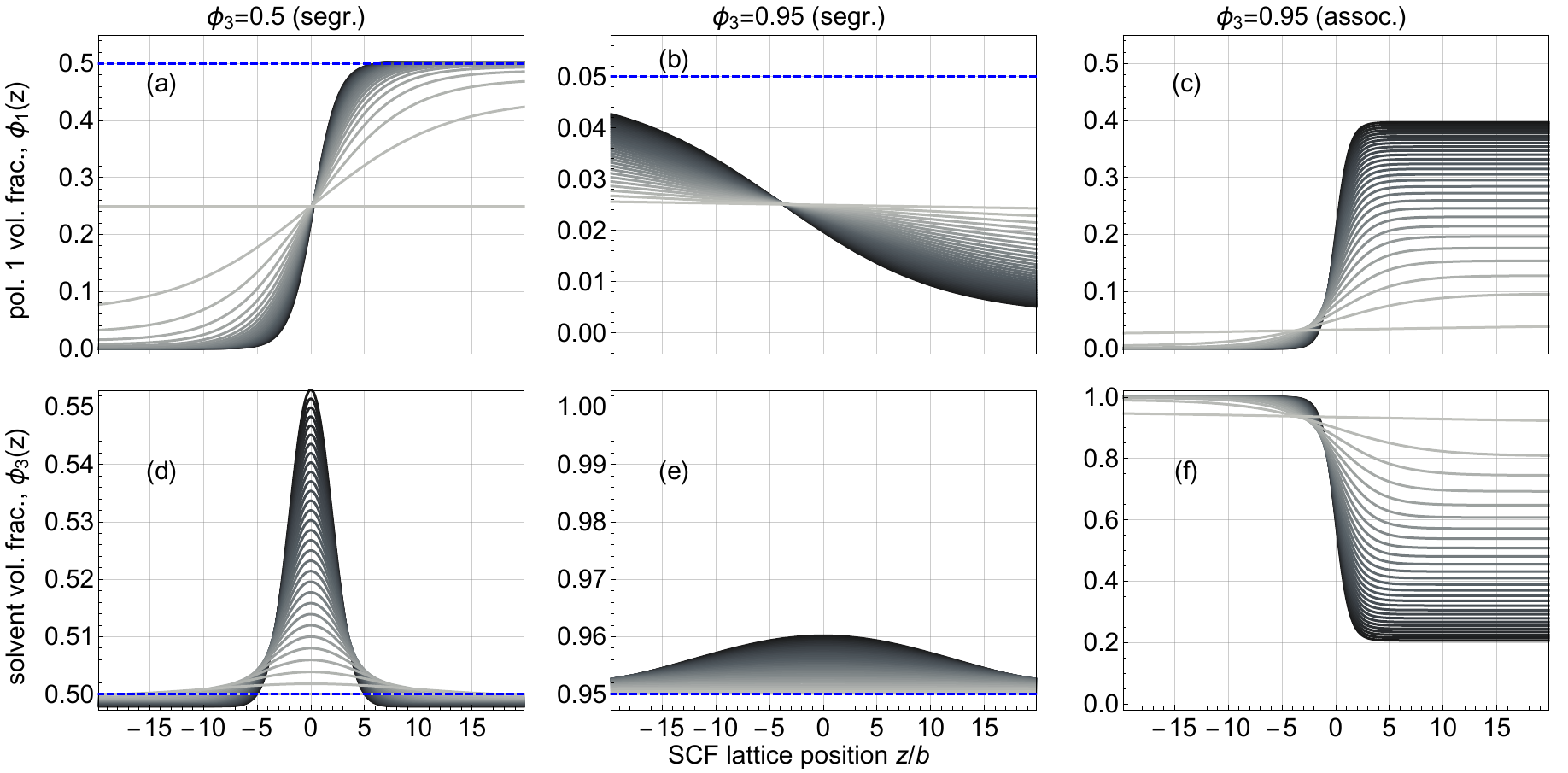}
    \caption{Similar to Fig. 5 in the main text, but for $\chi_{13}=0$. In the segregative scenario (a,b,d,e), the polymer--polymer interaction parameter $\chi_{12}$ decreases from $0.3$ to the corresponding critical value: for $\phi_3 = 0.5$ (segr.), $\chi_{12} = \{0.3, 0.296,...,0.032,0.028\}$ for $\phi_3 = 0.95$ (segr.), $\chi_{12} = \{0.3, 0.297,...,0.204,0.201\}$ ; (black to light gray as approaching the CP). In the association case (c,f) $\phi_3 = 0.95$ (assoc.), $\chi_{12} = \{-3.5,-3.46,-3.42,...,-2.02\}$ (approaching the CP from black to light gray). 
    }
    \label{fig:SCFprofs2}
\end{figure}

\clearpage

\subsection*{(Unscaled) Interfacial properties}
In the main text, we presented normalized interfacial tension and width values. Here we plot the raw values from the SCF computations.
\begin{figure}[htb!]
 \centering
    \includegraphics[width=0.99\textwidth]{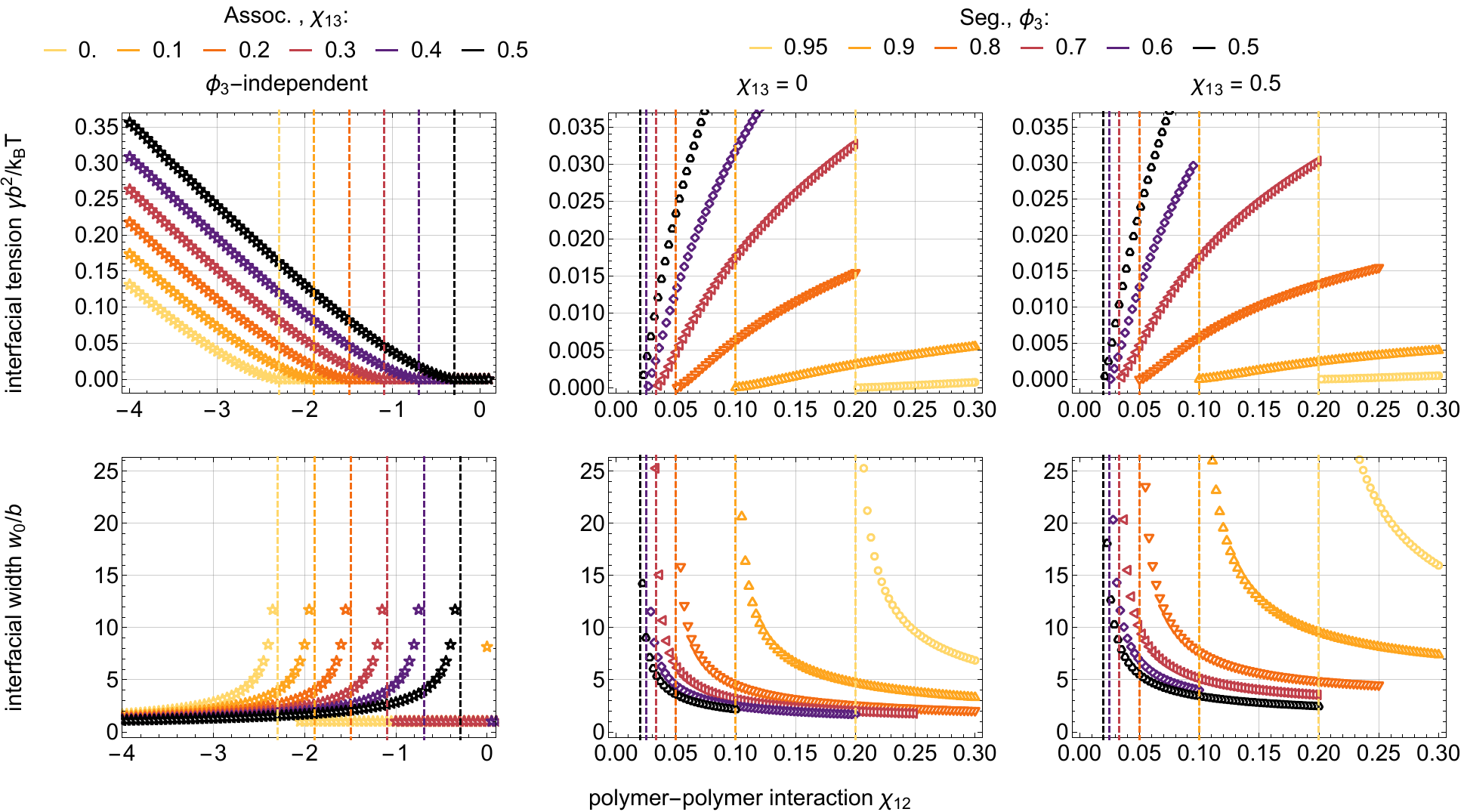}
    \caption{Interfacial tension $\gamma$ (a) and interfacial width $w_0$ from numerical SCF lattice computations (data) for both ALLPS and SLLPS as indicated in terms of the polymer--polymer interaction $\chi_{12}$. Dashed vertical lines indicate the critical point of the corresponding scenarios.}
    \label{fig:GridGammaAndw_SI}
\end{figure}

\end{document}